\documentclass[11pt]{article}
\usepackage{amsfonts}
\usepackage{amsmath}
\usepackage{amssymb}
\usepackage{indentfirst}
\usepackage{graphicx}
\usepackage[numbers,sort&compress]{natbib}

\begin{document}

\title{Generalized Pure Lovelock Gravity}
\author{Patrick Concha$^{1}$\thanks{%
patrick.concha@pucv.cl}, Evelyn Rodr\'{\i}guez$^{2}$\thanks{%
evelyn.rodriguez@edu.uai.cl}, \\
{\small $^{1}$\textit{Instituto de F\'{\i}sica, Pontificia Universidad Cat%
\'{o}lica de Valpara\'{\i}so,}}\\
\ {\small Casilla 4059, Valparaiso, Chile.}\\
{\small $^{2}$\textit{Departamento de Ciencias, Facultad de Artes Liberales,}%
}\\
{\small \textit{Universidad Adolfo Ib\'{a}\~{n}ez,}}\\
{\small Av. Padre Hurtado 750, Vi\~{n}a del Mar, Chile}\\
}
\maketitle

\begin{abstract}
We present a generalization of the n-dimensional (pure) Lovelock Gravity
theory based on an enlarged Lorentz symmetry. In particular, we propose an
alternative way to introduce a cosmological term. Interestingly, we show
that the usual pure Lovelock gravity is recovered in a matter-free
configuration. The five and six-dimensional cases are explicitly studied.
\end{abstract}

\bigskip

\bigskip

\section{Introduction}

The most natural generalization of General Relativity (GR) in $d$ dimensions
satisfying the criteria of general covariance and leading to second order
field equations for the metric is given by the Lanczos-Lovelock (LL) gravity
theory \cite{Lovelock:1971yv,Lanczos}. In the differential forms language
the LL action can be written as the most general $d$-form invariant under
local Lorentz transformations, constructed out of the spin connection $%
\omega ^{ab}$, the vielbein $e^{a}$ and their exterior derivatives \cite%
{Zumino, TeitZ},%
\begin{equation}
S_{LL}=\int \sum_{p=0}^{\left[ d/2\right] }\alpha _{p}\epsilon
_{a_{1}a_{2}\cdots a_{d}}R^{a_{1}a_{2}}\cdots
R^{a_{2p-1}a_{2p}}e^{a_{2p+1}}\cdots e^{a_{d}}\,,
\end{equation}%
where $R^{ab}=d\omega ^{ab}+\omega _{\text{ }c}^{a}\omega ^{cb}$ is the
Lorentz curvature two-form and the $\alpha _{p}$ coefficients are not fixed
from first principles.

Different numbers of degrees of freedom emerge depending on the value of the
arbitrary coefficients. In particular, the higher curvature terms can
produce degenerate sectors with no degrees of freedom. Such degeneracy can
be avoided with particular choices of the $\alpha _{p}$ constants. In
particular, as explained in ref.~\cite{DPP}, there are mainly two ways to
avoid degenerate sectors. One of them consist in restrict the theory to have
a unique degenerate vacuum which leads to a family of gravity theories \cite%
{CTZ} labeled by the integer $k$ which represents the highest power of
curvature. Interestingly, the $\alpha _{p}$ constants can be fixed requiring
that the theory has the maximum possible number of degrees of freedom \cite%
{TZ}. Then the LL Lagrangian is a Chern-Simons (CS) form \cite{Cham1, Cham2,
Zan} in odd dimensions which is gauge invariant under the $(A)dS$ symmetry.
In even dimensions, the LL Lagrangian can be written as a Born-Infeld (BI)
gravity Lagrangian \cite{BTZ, DG} which is locally invariant under a Lorentz
subalgebra.

Another way of fixing the $\alpha _{p}$ constants avoiding degeneracy is to
demand that there is non-degenerate vacuum. Such requirement leads to the
pure Lovelock (PL) theory \cite{DPP, CO} which consists only in two terms of
the full Lovelock Lagrangian,%
\begin{equation}
S_{PL}=\int \left( \alpha _{0}\mathcal{L}_{0}+\alpha _{p}\mathcal{L}%
_{p}\right) \,,
\end{equation}%
with%
\begin{eqnarray}
\mathcal{L}_{0} &=&\epsilon _{a_{1}\dots a_{d}}e^{a_{1}}\cdots e^{a_{d}}\,,
\\
\mathcal{L}_{p} &=&\epsilon _{a_{1}\dots a_{d}}R^{a_{1}a_{2}}\cdots
R^{a_{2p-1}a_{2p}}e^{a_{2p+1}}\cdots e^{a_{d}}\,.
\end{eqnarray}%
The coefficients are fixed in terms of the gravitational constant $\kappa $
and the cosmological constant $\Lambda $,%
\begin{eqnarray}
\alpha _{0} &=&-\frac{2\Lambda \kappa }{d!}=-\frac{\left( \mp 1\right)
^{p}\kappa }{d\left( d-2p-1\right) !\ell ^{2p}}\,, \\
\alpha _{p} &=&\frac{\kappa }{\left( d-2p\right) !}\,.
\end{eqnarray}%
With this particular choice, the PL theory has a unique nondegenerate $(A)dS$
vacuum in odd dimensions and admits non-degenerate vacua in even dimensions.
Additionally, the black holes (BH) solutions of the PL theory behave
asymptotically like the $AdS$-Schwarzschild ones \cite{DPP, DPP2}. Of
particular interest are the BH solutions of the maximal pure Lovelock since
the thermodynamical parameters are universal in terms of horizon radius.
Recently, a Hamiltonian analysis has shown that the maximum possible number
of degrees of freedom of the PL case is the same as in the Einstein-Hilbert
(EH) gravity \cite{DDMM}.

The supersymmetric version of the general LL theory is unknown, except in
the EH case and in odd dimensions when the LL action can be seen as a CS
supergravity action for the $AdS$ superalgebra. The construction of a super
LL gravity action, and in particular for a super PL, remains a difficult
task. Indeed, there is no clarity in which terms should be considered in the
action in order to guarantee the supersymmetric invariance of the theory. A
discussion about a five-dimensional supergravity action for the EH term
coupled to a Lovelock term can be found in ref.~\cite{DF}. \ More recently,
the authors of refs.~\cite{CDIMR, CMR} suggested that the supersymmetric
version of a PL theory could emerge as a particular limit of a supergravity
theory. The procedure that could be used is not new and has already been
used to relate GR with different gravity theories \cite{EHTZ, GRCS, CPRS1,
CPRS2, CPRS3}.

As shown in refs.~\cite{CDIMR, CMR}, the PL Lagrangian can be recovered as a
particular limit of CS and BI like Lagrangians constructed out the $%
\mathfrak{C}_{k}$ family \cite{CDMR}. Although the procedure presented in
\cite{CDIMR, CMR} can be reproduced in any spacetime dimension $d$, the
obtention of the PL action requires a large amount of extra fields in higher
dimensions. Additionally in order to recover the PL dynamics, it is
necessary to impose additional restrictions on the fields.

Here we present a generalized pure Lovelock (GPL) gravity theory which leads
to the PL action and its dynamics in a matter-free configuration limit
without further considerations. In particular, the field content of the
theory is spacetime dimension independent. Interestingly, the GPL action
allows us to introduce alternatively a generalized cosmological term which
generalizes the result obtained in ref.~\cite{AKL} to higher dimensions.
Moreover, we show that the GPL gravity action corresponds to a particular
case of a more generalized Lovelock (GL) gravity theory. We also show that
there is a particular choice of the coefficients appearing in the GL action
leading in odd and even dimensions to a CS and BI like gravity, respectively.

\section{Generalized Lovelock gravity action}

A generalization of the Lanczos-Lovelock gravity action can be performed
enlarging the Lorentz symmetry. A Generalized Lovelock (GL) gravity action
can be written as the most general $d$-form invariant under local
Lorentz-like transformations, constructed with the spin connection $\omega
^{ab}$, the vielbein $e^{a}$, a Lorentz-like field $k^{ab}$ and their
exterior derivatives, without the Hodge dual,%
\begin{equation}
S_{GL}=\int \,\sum_{p=0}^{\left[ d/2\right] }\sum_{m=0}^{p}\alpha _{p}\binom{%
p}{p-m}L_{GL}^{\left( p\right) }\,,  \label{GL}
\end{equation}%
where%
\begin{equation}
L_{GL}^{\left( p\right) }=\epsilon _{a_{1}a_{2}\cdots
a_{d}}R^{a_{1}a_{2}}\cdots R^{a_{2m-1}a_{2m}}F^{a_{2m+1}a_{2m+2}}\cdots
F^{a_{2p-1}a_{2p}}e^{a_{2p+1}}\cdots e^{a_{d}}\,,
\end{equation}%
with%
\begin{eqnarray}
R^{ab} &=&d\omega ^{ab}+\omega _{\text{ }c}^{a}\omega ^{cb}\,,  \label{2fc1}
\\
F^{ab} &=&dk^{ab}+\omega _{\text{ }c}^{a}k^{cb}-\omega _{\text{ }%
c}^{b}k^{ca}+k_{\text{ }c}^{a}k^{cb}\,.  \label{2fc2}
\end{eqnarray}%
Here the $\alpha _{p}$ coefficients are arbitrary constants which are not
fixed from first principles. The possible permutations of the curvature $2$%
-forms $R^{ab}$ and $F^{ab}$ appearing in the action are reflected in the
coefficients $\binom{p}{p-m}$.

Let us note that for $p=0$ the Lagrangian reproduces the cosmological
constant term, while for $p=1$ the GL Lagrangian is the Einstein-Hilbert
term plus the coupling of $F^{ab}$ with the vielbeins. This is an important
difference with the usual Lovelock-Cartan gravity theory which contains
General Relativity as a particular case. However, a matter-free
configuration ($k^{ab}=0$) allows to recover the usual Lovelock gravity
action. In particular, the Einstein-Hilbert term is recovered for $p=1$ and $%
k^{ab}=0$.

The dynamical content is obtained considering the variation of the GL action
with respect to $\left( e^{a},\omega ^{ab},k^{ab}\right) $,%
\begin{equation}
\delta S_{GL}=\int \delta e^{a}\mathcal{E}_{a}+\delta \omega ^{ab}\mathcal{E}%
_{ab}+\delta k^{ab}\mathcal{K}_{ab}=0\,,
\end{equation}%
modulo boundary terms. The field equations are given by%
\begin{eqnarray}
\mathcal{E}_{a} &=&\sum_{p=0}^{\left[ \frac{d-1}{2}\right]
}\sum_{m=0}^{p}\alpha _{p}\left( d-2p\right) \mathcal{E}_{a}^{p}=0\,, \\
\mathcal{E}_{ab} &=&\sum_{p=1}^{\left[ \frac{d-1}{2}\right]
}\sum_{m=0}^{p}\alpha _{p}m\left( d-2p\right) \mathcal{E}_{ab}^{p}=0\,,
\label{cd} \\
\mathcal{K}_{ab} &=&\sum_{p=1}^{\left[ \frac{d-1}{2}\right]
}\sum_{m=0}^{p}\alpha _{p}\left( p-m\right) \left( d-2p\right) \mathcal{E}%
_{ab}^{p}=0\,,  \label{cd2}
\end{eqnarray}%
where%
\begin{eqnarray}
\mathcal{E}_{a}^{p} &\equiv &\binom{p}{p-m}\epsilon _{ab_{1}\cdots
b_{d-1}}R^{b_{1}b_{2}}\cdots R^{b_{2m-1}b_{2m}}F^{b_{2m+1}b_{2m+2}}\cdots
F^{b_{2p-1}b_{2p}}e^{b_{2p+1}}\cdots e^{b_{d-1}}, \\
\mathcal{E}_{ab}^{p} &\equiv &\,\binom{p}{p-m}\epsilon _{aba_{3}\cdots
a_{d}}R^{a_{3}a_{4}}\cdots R^{a_{2m-1}a_{2m}}F^{a_{2m+1}a_{2m+2}}\cdots
F^{a_{2p-1}a_{2p}}R^{a_{2p+1}}e^{a_{2p+2}}\cdots e^{a_{d}}\,.
\end{eqnarray}%
Here $R^{a}=D_{\omega }e^{a}+k_{\text{ }b}^{a}e^{b}$ with $D_{\omega
}=d+\omega $ the Lorentz covariant exterior derivative. In particular, the
Bianchi identities $D_{\omega }R^{ab}=0$ and $D_{\omega +k}F^{ab}+k_{\text{ }%
c}^{a}R^{cb}-k_{\text{ }c}^{b}R^{ca}=0$ along with $d^{2}=0,$ assure that
the field equations involve only first derivatives of $e^{a},$ $\omega ^{ab}$
and $k^{ab}$.

Let us note that the variation of the action under the spin connection $%
\omega ^{ab}$ and the Lorentz-like field $k^{ab}$ imply the same field
equation and then the $\left( d-1\right) $-forms $\mathcal{E}_{ab}$ and $%
\mathcal{K}_{ab}$ coincide. \ On the other hand, analogously to the usual
Lovelock Cartan gravity, the $\left( d-1\right) $-form $\mathcal{E}_{a}$ is
independent of the $\left( d-1\right) $-forms $\mathcal{E}_{ab}$.

Furthermore, using the Bianchi identities for the curvature $2$-forms one
can show that the following relation holds%
\begin{equation}
D\mathcal{E}_{a}^{p}=\left( d-1-2p\right) e^{b}\mathcal{E}_{ba}^{p+1}\,,
\end{equation}%
for $0\leq p\leq \left[ \frac{d-1}{2}\right] $. Then, as in refs.~\cite{TZ,
CPRS3} we have that the previous identity leads to%
\begin{equation}
D\mathcal{E}_{a}=\sum_{p=1}^{\left[ \frac{d+1}{2}\right] }\sum_{m=0}^{p}%
\alpha _{p-1}\left( d+2-2p\right) \left( d+1-2p\right) e^{b}\mathcal{E}%
_{ba}^{p}\,,  \label{eq1}
\end{equation}%
which by consistency with $\mathcal{E}_{a}=0$ must also be zero. Besides, we
can see that the following product%
\begin{equation}
e^{b}\mathcal{E}_{ba}=\sum_{p=1}^{\left[ \frac{d-1}{2}\right]
}\sum_{m=0}^{p}\alpha _{p}m\left( d-2p\right) e^{b}\mathcal{E}_{ba}^{p}
\label{eq2}
\end{equation}%
must also vanish by consistency with $\mathcal{E}_{ab}=0$. One can easily
check that the same result apply for the product $e^{b}\mathcal{K}_{ba}$.

Thus, fixing the $\alpha _{p}$ coefficients could lead to different numbers
of degrees of freedom depending on additional constraints of the form $e^{b}%
\mathcal{E}_{ba}^{p}=0$. Following the same arguments of ref.~\cite{TZ},
there is a particular choice in odd dimensions allowing to avoid additional
restrictions and such that $\mathcal{E}_{a}~$and $\mathcal{E}_{ab}$ (or $%
\mathcal{K}_{ab}$) are independent.

\subsection{Chern-Simons gravity and $\mathfrak{C}_{4}$ algebra}

The odd-dimensional GL action (\ref{GL}) reproduces a Chern-Simons action
for a particular Lie algebra, known as the $\mathfrak{C}_{4}$ algebra%
\footnote{Also known as Poincar\'{e}
semi-simple extended algebra.}
\cite{Sorokas, DFIMRSV, SS, CDMR}, when the $\alpha _{p}$'s are fixed to the
following values:%
\begin{eqnarray}
\alpha _{0} &=&\frac{\kappa }{d\ell ^{d}}\,, \\
\alpha _{p} &=&\alpha _{0}\frac{\left( 2n-1\right) \left( 2\gamma \right)
^{p}}{\left( 2n-2p-1\right) }\binom{n-1}{p}\,,
\end{eqnarray}%
where $\gamma $ is related to the cosmological constant,%
\begin{equation}
\gamma =-sgn\left( \Lambda \right) \frac{\ell ^{2}}{2}\,,
\end{equation}%
and $\ell $ is a length parameter related to the cosmological constant.

The $d=2n-1$ CS Lagrangian is given by%
\begin{eqnarray}
L_{CS}^{\mathfrak{C}_{4}} &=&\kappa \epsilon _{a_{1}a_{2}\cdots
a_{2n-1}}\sum_{p=0}^{n-1}\sum_{m=0}^{p}\ell ^{2\left( p-n\right) +1}c_{p}%
\binom{p}{p-m}  \notag \\
&&\times R^{a_{1}a_{2}}\cdots R^{a_{2m-1}a_{2m}}F^{a_{2m+1}a_{2m+2}}\cdots
F^{a_{2p-1}a_{2p}}e^{a_{2p+1}}\cdots e^{a_{2n-1}}\,,
\end{eqnarray}%
with%
\begin{equation}
c_{p}=\frac{1}{2\left( n-p\right) -1}\binom{n-1}{p}\,.
\end{equation}%
Let us note that the $d=3$ CS form reproduces the generalized CS gravity
theory presented in \cite{DFIMRSV} applying an appropriate change of basis
where the $AdS\oplus Lorentz$ structure is manifested. The black hole
solution of the three-dimensional CS gravity theory based on this symmetry
have been recently studied in \cite{HR}.

It is important to clarify that, unlike the GL Lagrangian, the CS $\left(
2n-1\right) $-form is invariant not only under Lorentz-like transformation,%
\begin{eqnarray}
\delta e^{a} &=&\,e^{b}\rho _{b}^{\text{ }a}+e^{b}\lambda _{b}^{\text{ }a},
\\
\delta \omega ^{ab} &=&D_{\omega }\rho ^{ab}\,, \\
\delta k^{ab} &=&\,D_{\omega }\lambda ^{ab}+k_{\text{ }c}^{a}\rho ^{cb}-k_{%
\text{ }c}^{b}\rho ^{ca}+k_{\text{ }c\text{ }}^{a}\lambda ^{cb}-k_{\text{ }%
c}^{b}\lambda ^{ca},
\end{eqnarray}%
but also under a local $\mathfrak{C}_{4}$ boost%
\begin{eqnarray}
\delta e^{a} &=&D_{\omega }\rho ^{a}\,+k_{\text{ }b}^{a}\rho ^{a}, \\
\delta \omega ^{ab} &=&0\,, \\
\delta k^{ab} &=&\rho ^{a}e^{b}-\rho ^{b}e^{a}\,,
\end{eqnarray}%
where the $\mathfrak{C}_{4}$ gauge parameter is given by%
\begin{equation}
\rho =\frac{1}{2}\rho ^{ab}J_{ab}+\frac{1}{2}\lambda ^{ab}Z_{ab}+\frac{1}{%
\ell }\rho ^{a}P_{a}\,.
\end{equation}%
In particular, the generators of the $\mathfrak{C}_{4}$ algebra satisfy the
following commutation relations:%
\begin{align}
\left[ J_{ab},J_{cd}\right] & =\eta _{bc}J_{ad}-\eta _{ac}J_{bd}-\eta
_{bd}J_{ac}+\eta _{ad}J_{bc}\,,  \label{c4e1} \\
\left[ J_{ab},Z_{cd}\right] & =\eta _{bc}Z_{ad}-\eta _{ac}Z_{bd}-\eta
_{bd}Z_{ac}+\eta _{ad}Z_{bc}\,, \\
\left[ Z_{ab},Z_{cd}\right] & =\eta _{bc}Z_{ad}-\eta _{ac}Z_{bd}-\eta
_{bd}Z_{ac}+\eta _{ad}Z_{bc}\,,  \label{c4e3} \\
\left[ J_{ab},P_{c}\right] & =\eta _{bc}P_{a}-\eta _{ac}P_{b}\,,\text{ \ \ \
\ }\left[ P_{a},P_{b}\right] =Z_{ab}\,, \\
\left[ Z_{ab},P_{c}\right] & =\eta _{bc}P_{a}-\eta _{ac}P_{b}\,,
\label{c4e5}
\end{align}%
Such symmetry can be obtained as a deformation of the Maxwell symmetry and
belongs to a generalized family of Lie algebras denoted by $\mathfrak{C}_{k}$
\cite{CDMR}. At the fermionic level, a recent application has been developed
in the three-dimensional CS context where a $\left( p,q\right) $ $AdS$%
-Lorentz supergravity model was presented \cite{CFR}.

\subsection{Born-Infeld gravity and Lorentz-like symmetry}

The even-dimensional case requires an alternative approach since eqs. (\ref%
{eq1}) and (\ref{eq2}) have not the same number of terms as in the
odd-dimensional case.

Following the same procedure introduced in refs.~\cite{TZ, CPRS3}, one can
note that there is a particular choice of the $\alpha _{p}$'s
\begin{eqnarray}
\alpha _{0} &=&\frac{\kappa }{d\ell ^{d}}\,, \\
\alpha _{p} &=&\alpha _{0}\left( 2\gamma \right) ^{p}\binom{n}{p}\,,
\end{eqnarray}%
which reproduces a Born-Infeld (BI) like gravity action with $0\leq p\leq n$%
. As in the CS case, $\gamma $ is related to the cosmological constant $%
\gamma =-sgn\left( \Lambda \right) \frac{\ell ^{2}}{2}$. With these
coefficients the GL Lagrangian takes the BI-like form%
\begin{eqnarray}
L_{BI}^{\mathcal{L}_{\mathfrak{C}_{4}}} &=&\kappa \epsilon
_{a_{1}a_{2}\cdots a_{2n}}\sum_{p=0}^{n}\sum_{m=0}^{p}\frac{\ell ^{2p-2n}}{2n%
}\binom{n}{p}\binom{p}{p-m}  \notag \\
&&\times R^{a_{1}a_{2}}\cdots R^{a_{2m-1}a_{2m}}F^{a_{2m+1}a_{2m+2}}\cdots
F^{a_{2p-1}a_{2p}}e^{a_{2p+1}}\cdots e^{a_{2n}}\,.  \label{lbi}
\end{eqnarray}%
In particular, the Lagrangian can be rewritten in a reduced form,%
\begin{equation}
L_{BI}^{\mathcal{L}_{\mathfrak{C}_{4}}}=\frac{\kappa }{2n}\epsilon
_{a_{1}a_{2}\cdots a_{2n}}\bar{F}^{a_{1}a_{2}}\dots \bar{F}%
^{a_{2n-1}a_{2n}}\,,  \label{lbi2}
\end{equation}%
where%
\begin{equation}
\bar{F}^{ab}=R^{ab}+F^{ab}+\frac{1}{\ell ^{2}}e^{a}e^{b}\,,
\end{equation}%
is the $\mathfrak{C}_{4}$ curvature. With this form, the Lagrangian
corresponds to the Pfaffian of the $2$-form $\bar{F}^{ab}$ and can be
rewritten similarly to the Born-Infeld electrodynamics Lagrangian,%
\begin{equation}
L_{BI}^{\mathcal{L}_{\mathfrak{C}_{4}}}=2^{n-1}\left( n-1\right) !\sqrt{\det
\left( R^{ab}+F^{ab}+\frac{1}{\ell ^{2}}e^{a}e^{b}\right) }\,.
\end{equation}

It is important to emphasize that the BI-like gravity Lagrangian is only
off-shell invariant under a Lorentz-like subalgebra\ $\mathcal{L}_{\mathfrak{%
C}_{4}}$ of the $\mathfrak{C}_{4}$ algebra. This can be clarified through
the Levi-Civita symbol $\epsilon _{a_{1}a_{2}\cdots a_{2n}}$ in (\ref{lbi2})
which consists in the only non-vanishing component of the Lorentz-like
invariant tensor of rank $n$, namely%
\begin{equation}
\left\langle \bar{Z}_{a_{1}a_{2}}\cdots \bar{Z}_{a_{2n-1}a_{2n}}\right%
\rangle =\frac{2^{n-1}}{n}\epsilon _{a_{1}a_{2}\cdots a_{2n}}\,.
\end{equation}%
Here $\bar{Z}_{ab}=J_{ab}+Z_{ab}$ satisfy the $\mathfrak{C}_{4}$ commutation
relations (\ref{c4e1})-(\ref{c4e3}). Such choice of the invariant tensor
breaks the full $\mathfrak{C}_{4}$ symmetry to its Lorentz-like subgroup $%
\mathcal{L}_{\mathfrak{C}_{4}}$.

Recently, diverse BI-like gravity theories have been constructed with
different purposes \cite{CPRS1, CPRS2, CMR}. At the supersymmetric level,
similar constructions have been done based on the MacDowell-Mansouri
formalism \cite{MM, TPN, CR2, CRS, AD, CIRR, PR}.

\section{Generalized Pure Lovelock gravity action}

An alternative way of fixing the $\alpha _{p}$'s can be implemented such
that a generalized pure Lovelock (GPL) gravity action can be constructed.
The new coefficients are fixed in terms of the gravitational constant $%
\kappa $ and the cosmological constant $\Lambda $ in the same way as in
refs.~\cite{CO, DPP} ,%
\begin{eqnarray}
\alpha _{0} &=&-\frac{2\Lambda \kappa }{d!}=-\frac{\left( \mp 1\right)
^{p}\kappa }{d\left( d-2p-1\right) !\ell ^{2p}}\,,  \label{alpha0} \\
\alpha _{p} &=&\frac{\kappa }{\left( d-2p\right) !}\,.  \label{alphap}
\end{eqnarray}%
Such generalization contains only two term of the full GL series (given by
eq. (\ref{GL})),%
\begin{equation}
S_{GPL}=\int \left( \alpha _{0}\mathcal{L}_{0}+\alpha _{p}\mathcal{L}%
_{p}\right) \,,  \label{GPL}
\end{equation}%
where%
\begin{eqnarray}
\mathcal{L}_{0} &=&\epsilon _{a_{1}a_{2}\dots a_{d}}e^{a_{1}}e^{a_{2}}\cdots
e^{a_{d}}\,, \\
\mathcal{L}_{p} &=&\sum_{m=0}^{p}\binom{p}{p-m}\epsilon _{a_{1}a_{2}\dots
a_{d}}R^{a_{1}a_{2}}\cdots R^{a_{2m-1}a_{2m}}F^{a_{2m+1}a_{2m+2}}\cdots
F^{a_{2p-1}a_{2p}}e^{a_{2p+1}}\cdots e^{a_{d}}\,.
\end{eqnarray}

Interestingly, the $p$-order term in the curvature $2$-forms reproduces a
generalized cosmological term in any spacetime dimension $d$. This
particular case of the generalized Lovelock series has been first introduced
in four dimensions in refs.~\cite{AKL, CRS}. Subsequently in ref.~\cite{CMR}%
, a generalized cosmological constant term has been obtained in even
dimensions as a particular configuration limit of a BI-like gravity theory.
Thus, the generalized pure Lovelock action presented here, through the
Lorentz-like gauge field $k^{ab}$,$\,\ $allows us to introduce alternatively
a cosmological term in arbitrary dimensions.

Let us note that the Einstein-Hilbert term appears only in the case $p=1$
along with the following term:%
\begin{equation}
\epsilon _{a_{1}a_{2}a_{3}\dots a_{d}}F^{a_{1}a_{2}}e^{a_{3}}\cdots
e^{a_{d}}\,.
\end{equation}

On the other hand, topological densities are obtained in even dimensions for
$p=d/2$,%
\begin{equation}
\mathcal{L}_{d/2}=\sum_{m=0}^{p}\binom{d/2}{d/2-m}\epsilon _{a_{1}a_{2}\dots
a_{d}}R^{a_{1}a_{2}}\cdots R^{a_{2m-1}a_{2m}}F^{a_{2m+1}a_{2m+2}}\cdots
F^{a_{d-1}a_{d}}\,.
\end{equation}%
These terms are related to Euler type characteristic classes. Then, we have
that for $1\leq p\leq \frac{d-2}{2}$ the even-dimensional GPL gravity action
reproduces truly dynamical actions.

Interestingly, let us note that in a matter-free configuration $k^{ab}=0$
the GPL action (\ref{GPL}) reduces to the pure Lovelock action,%
\begin{equation}
S_{PL}=\int \left( \alpha _{0}\mathcal{L}_{0}+\alpha _{p}\mathcal{L}%
_{p}\right) \,,
\end{equation}%
with%
\begin{eqnarray}
\mathcal{L}_{0} &=&\epsilon _{a_{1}a_{2}\dots a_{d}}e^{a_{1}}e^{a_{2}}\cdots
e^{a_{d}}\,, \\
\mathcal{L}_{p} &=&\epsilon _{a_{1}a_{2}\dots a_{d}}R^{a_{1}a_{2}}\cdots
R^{a_{2p-1}a_{2p}}e^{a_{2p+1}}\cdots e^{a_{d}}\,.
\end{eqnarray}%
Obtaining the PL gravity action in a particular matter-free configuration
limit is not new and has already been presented in refs.~\cite{CDIMR, CMR}.
Nevertheless, the techniques considered in \cite{CDIMR, CMR} require an
excessive amount of new extra fields as the spacetime dimension grows. In
our case, the field content of the theory does not depend of the spacetime
dimension $d$. Furthermore, one can show that the PL dynamics \cite{CO, DPP,
DPP2} can also be reproduced in a matter-free configuration leading
appropriately to the pure Lovelock gravity theory. In fact considering $%
k^{ab}=0$, we have%
\begin{eqnarray}
&&\alpha _{0}\epsilon _{a_{1}a_{2}\cdots a_{d}}e^{a_{1}}e^{a_{2}}\cdots
e^{a_{d-1}}+\alpha _{p}\epsilon _{a_{1}a_{2}\cdots
a_{d}}R^{a_{1}a_{2}}\cdots R^{a_{2p-1}a_{2p}}e^{a_{2p+1}}\cdots
e^{a_{d-1}}=0, \\
&&\alpha _{p}\epsilon _{a_{1}a_{2}\cdots a_{d}}R^{a_{3}a_{4}}\cdots
R^{a_{2p-1}a_{2p}}T^{a_{2p+1}}e^{a_{2p+2}}\cdots e^{a_{d}}\,=0,
\end{eqnarray}%
where $T^{a}=D_{\omega }e^{a}$. Unlike the procedure presented in previous
works, the truly PL dynamics is recovered here without imposing any
identification on the fields.

One could obtain the same result considering $k^{ab}$ as the true
spin-connection and $\omega ^{ab}$ as the new extra-field, but it is
straightforward to see that the curvatures (\ref{2fc1})-(\ref{2fc2})
reproduces the usual Lorentz curvature only when $\omega ^{ab}$ is
identified as the true spin-connection one-form.

Thus, we have obtained the PL theory considering not only an appropriate
limit in the GPL action (\ref{GPL}) but also a right dynamical limit without
any identification of the fields.

\subsection{The five-dimensional case}

The five-dimensional generalized pure Lovelock gravity reproduces two
diverse actions depending on the value of $p$. Indeed, for $p=1$ the GPL
action reduces to%
\begin{equation}
S_{GPL}^{p=1}=\int_{M_{5}}\epsilon _{abcde}\left[ \alpha
_{0}e^{a}e^{b}e^{c}e^{d}e^{e}+\alpha _{1}\left(
R^{ab}e^{c}e^{d}e^{e}+F^{ab}e^{c}e^{d}e^{e}\right) \right] \,.
\end{equation}%
Here $\alpha _{0}$ and $\alpha _{1}$ are given by eqs. (\ref{alpha0})-(\ref%
{alphap}). The $p=1$ GPL action can be seen as the coupling of a generalized
cosmological term $\mathcal{L}_{\tilde{\Lambda}}$ to the Einstein-Hilbert
term,%
\begin{equation}
S_{GPL}^{p=1}=\int_{M_{5}}\mathcal{L}_{\tilde{\Lambda}}+\mathcal{L}_{EH}\,,
\end{equation}%
where%
\begin{equation}
\mathcal{L}_{\tilde{\Lambda}}=\alpha _{0}\epsilon
_{abcde}e^{a}e^{b}e^{c}e^{d}e^{e}+\alpha _{1}\epsilon _{abcde}\left(
D_{\omega }k^{ab}e^{c}e^{d}e^{e}+k_{\text{ }f}^{a}k^{fb}e^{c}e^{d}e^{e}%
\right) \,.
\end{equation}%
One can see that the matter-free configuration limit ($k^{ab}=0$) leads to
the $p=1$ pure Lovelock action%
\begin{equation}
S_{PL}^{p=1}=\int_{M_{5}}\epsilon _{abcde}\left( \alpha
_{0}e^{a}e^{b}e^{c}e^{d}e^{e}+\alpha _{1}R^{ab}e^{c}e^{d}e^{e}\right) \,,
\end{equation}%
where the $\alpha _{p}$ coefficients are identical to the PL ones. Let us
note that the $p=1$ PL action corresponds to the standard General Relativity
action in presence of a cosmological constant term. The obtention of GR in a
matter-free configuration limit of the GPL theory is a desirable feature in
order to generalize gravity since it should satisfy the correspondence
principle. Furthermore, in a matter-free configuration, the field equations
read%
\begin{eqnarray}
\epsilon _{abcde}\left( \alpha _{0}e^{a}e^{b}e^{c}e^{d}+\alpha
_{1}R^{ab}e^{c}e^{d}\right) \delta e^{e} &=&0\,, \\
\epsilon _{abcde}\left( \alpha _{1}T^{c}e^{d}e^{e}\right) \delta \omega
^{ab} &=&0\,, \\
\epsilon _{abcde}\left( \alpha _{1}T^{c}e^{d}e^{e}\right) \delta k^{ab}
&=&0\,,
\end{eqnarray}%
which correspond to the appropriate $p=1$ PL dynamics described in refs.~%
\cite{CO, DPP, DPP2}.

On the other hand, the $p=2$ case does not contain the EH term and the GPL
action is given by%
\begin{equation}
S_{GPL}^{p=2}=\int_{M_{5}}\epsilon _{abcde}\left[ \alpha
_{0}e^{a}e^{b}e^{c}e^{d}e^{e}+\alpha _{2}\left(
R^{ab}R^{cd}e^{e}+R^{ab}F^{cd}e^{e}+F^{ab}F^{cd}e^{e}\right) \right] \,,
\end{equation}%
which can be seen as the coupling of a generalized cosmological term $%
\mathcal{L}_{\hat{\Lambda}}$ to an Lanczos-Lovelock term $\mathcal{L}_{LL}$,%
\begin{equation}
S_{GPL}^{p=2}=\int_{M_{5}}\mathcal{L}_{\hat{\Lambda}}+\mathcal{L}_{LL}\,.
\end{equation}%
The $\mathcal{L}_{\hat{\Lambda}}$ term includes the usual cosmological term
plus additional terms depending on the Lorentz-like field $k^{ab}$,%
\begin{equation}
\mathcal{L}_{\hat{\Lambda}}=\alpha _{0}\epsilon
_{abcde}e^{a}e^{b}e^{c}e^{d}e^{e}+\alpha _{1}\epsilon _{abcde}\left(
R^{ab}Dk^{cd}e^{e}+Dk^{ab}Dk^{cd}e^{e}\right) \,,
\end{equation}%
with $D=d+\omega +k$.

As in the $p=1$ case, the $p=2$ PL theory is recovered in a matter-free
configuration limit,%
\begin{equation}
S_{PL}^{p=2}=\int_{M_{5}}\epsilon _{abcde}\left[ \alpha
_{0}e^{a}e^{b}e^{c}e^{d}e^{e}+\alpha _{2}R^{ab}R^{cd}e^{e}\right] \,,
\end{equation}%
while the field equations considering $k^{ab}=0$ reproduce the $p=2$ PL
dynamics,%
\begin{eqnarray}
\epsilon _{abcde}\left( \alpha _{0}e^{a}e^{b}e^{c}e^{d}+\alpha
_{1}R^{ab}R^{cd}\right) \delta e^{e} &=&0\,, \\
\epsilon _{abcde}\left( \alpha _{1}R^{cd}T^{e}\right) \delta \omega ^{ab}
&=&0\,, \\
\epsilon _{abcde}\left( \alpha _{1}R^{cd}T^{e}\right) \delta k^{ab} &=&0\,.
\end{eqnarray}

Let us note that the value and the sign of $\alpha _{0}$ is different for
any value of $p$ and thus $\alpha _{0}$ is distinct for $p=1$ and $p=2$. In
particular, for even value of $p$ the $\alpha _{0}$ coefficient has a
negative sign which makes the pure Lovelock theory to have a unique
nondegenerate $ds$ and $AdS$ vacuum \cite{DDMM}. Interestingly, no further
considerations on the $\alpha _{p}$ constants or in the fields must be
imposed in order to obtain appropriately the pure Lovelock theory. A similar
procedure has been considered in ref.~\cite{CDIMR} in order to recover the
five-dimensional PL theory. However, the obtention of the PL action and
dynamics in \cite{CDIMR} required the introduction not only of four new
extra-fields but also appropriate identifications on the extra-fields.

\subsection{The six-dimensional case}

The six-dimensional generalized pure Lovelock action also reproduces two
diverse gravity actions depending on the value of $p$. Each case describes
an alternative way to introduce a cosmological term. Let us note that only $%
p=1$ and $p=2$ reproduce non-trivial actions meanwhile the $p=3$ case does
not correspond to a GPL action since it leads to topological terms. For $%
d\geq 7$, a $p=3$ GPL action can be constructed.

The $p=1$ GPL action consists in the Einstein-Hilbert term plus a
six-dimensional generalized cosmological term,%
\begin{equation}
S_{GPL}^{p=1}=\int_{M_{6}}\epsilon _{abcdef}\left[ \alpha
_{0}e^{a}e^{b}e^{c}e^{d}e^{e}e^{f}+\alpha _{1}\left(
R^{ab}e^{c}e^{d}e^{e}e^{f}+F^{ab}e^{c}e^{d}e^{e}e^{f}\right) \right] \,.
\end{equation}%
The GPL action can be rewritten explicitly as%
\begin{equation}
S_{GPL}^{p=1}=\int_{M_{6}}\mathcal{L}_{\tilde{\Lambda}}+\mathcal{L}_{EH}\,,
\end{equation}%
where%
\begin{equation}
\mathcal{L}_{\tilde{\Lambda}}=\alpha _{0}\epsilon
_{abcdef}e^{a}e^{b}e^{c}e^{d}e^{e}e^{f}+\alpha _{1}\epsilon _{abcdef}\left(
D_{\omega }k^{ab}e^{c}e^{d}e^{e}e^{f}+k_{\text{ }%
f}^{a}k^{fb}e^{c}e^{d}e^{e}e^{f}\right) \,.
\end{equation}%
The $p=1$ GPL corresponds to one of the simplest generalizations of the GR
theory. Interestingly GR in presence of the cosmological constant, which
corresponds to the $p=1$ pure Lovelock action, emerges considering a
matter-free configuration limit ($k^{ab}=0$),%
\begin{equation}
S_{PL}^{p=1}=\int_{M_{6}}\epsilon _{abcdef}\left( \alpha
_{0}e^{a}e^{b}e^{c}e^{d}e^{e}e^{f}+\alpha
_{1}R^{ab}e^{c}e^{d}e^{e}e^{f}\right) \,.
\end{equation}%
The $p=1$ PL action always corresponds to the standard General Relativity
action in presence of a cosmological constant. Moreover, in a matter-free
configuration, the field equations read%
\begin{eqnarray}
\epsilon _{abcdef}\left( \alpha _{0}e^{a}e^{b}e^{c}e^{d}e^{e}+\alpha
_{1}R^{ab}e^{c}e^{d}e^{e}\right) \delta e^{f} &=&0\,, \\
\epsilon _{abcdef}\left( \alpha _{1}T^{c}e^{d}e^{e}e^{f}\right) \delta
\omega ^{ab} &=&0\,, \\
\epsilon _{abcdef}\left( \alpha _{1}T^{c}e^{d}e^{e}e^{f}\right) \delta
k^{ab} &=&0\,,
\end{eqnarray}%
which describe appropriately the $p=1$ PL dynamics.

On the other hand, the $p=2$ GPL action describes an alternative way to
introduce a cosmological term,%
\begin{equation}
S_{GPL}^{p=2}=\int_{M_{6}}\epsilon _{abcdef}\left[ \alpha
_{0}e^{a}e^{b}e^{c}e^{d}e^{e}e^{f}+\alpha _{2}\left(
R^{ab}R^{cd}e^{e}e^{f}+R^{ab}F^{cd}e^{e}e^{f}+F^{ab}F^{cd}e^{e}e^{f}\right) %
\right] \,,
\end{equation}%
which can be seen as the coupling of a generalized cosmological term $%
\mathcal{L}_{\hat{\Lambda}}$ to an Gauss-Bonnet term $\mathcal{L}_{GB}$,%
\begin{equation}
S_{GPL}^{p=2}=\int_{M_{6}}\mathcal{L}_{\hat{\Lambda}}+\mathcal{L}_{GB}\,.
\end{equation}%
Here%
\begin{equation}
\mathcal{L}_{\hat{\Lambda}}=\alpha _{0}\epsilon
_{abcde}e^{a}e^{b}e^{c}e^{d}e^{e}e^{f}+\alpha _{1}\epsilon _{abcdef}\left(
R^{ab}Dk^{cd}e^{e}e^{f}+Dk^{ab}Dk^{cd}e^{e}e^{f}\right) ,
\end{equation}%
with $D=d+\omega +k$.

Considering $k^{ab}=0$ we recover the six-dimensional $p=2$ PL theory,%
\begin{equation}
S_{PL}^{p=2}=\int_{M_{6}}\epsilon _{abcdef}\left[ \alpha
_{0}e^{a}e^{b}e^{c}e^{d}e^{e}e^{f}+\alpha _{2}R^{ab}R^{cd}e^{e}e^{f}\right]
\,,
\end{equation}%
while the field equations read in a matter-free configuration limit%
\begin{eqnarray}
\epsilon _{abcdef}\left( \alpha _{0}e^{a}e^{b}e^{c}e^{d}e^{f}+\alpha
_{1}R^{ab}R^{cd}e^{e}\right) \delta e^{f} &=&0\,, \\
\epsilon _{abcdef}\left( \alpha _{1}R^{cd}T^{e}e^{f}\right) \delta \omega
^{ab} &=&0\,, \\
\epsilon _{abcdef}\left( \alpha _{1}R^{cd}T^{e}e^{f}\right) \delta k^{ab}
&=&0\,.
\end{eqnarray}%
which correspond to the appropriate $p=1$ PL dynamics \cite{CO, DPP, DPP2}.

As in odd-dimensions, no further considerations have to be imposed in order
to obtain appropriately the pure Lovelock theory. A similar procedure has
been considered in ref.~\cite{CMR} in order to recover the even-dimensional
PL theory from a Born-Infeld like gravity theory. Nevertheless, the
obtention of the PL theory in ref.~\cite{CMR} requires much more conditions
and an excessive amount of extra fields.

\section{Discussion}

In the present work, we have presented a generalized Lovelock gravity theory
introducing an additional field which enlarge the symmetry to a Lorentz-like
algebra. Interestingly, a generalized pure Lovelock theory is obtained
fixing the $\alpha _{p}$ coefficients which consists only in two terms of
full generalized Lovelock action. The generalized PL action considered here
allows us to present an alternative way of introducing a generalized
cosmological term. Our result generalizes the four-dimensional case
presented in ref.~\cite{AKL} to arbitrary dimensions.

In addition, the usual pure Lovelock theory is recovered in a matter-free
configuration of the GPL theory. Unlike refs.~\cite{CDIMR, CMR},\ not only
the PL action \ is recovered but also the right PL dynamics is directly
obtained in the matter-free configuration limit. Such limit is considered
without imposing any identifications on the fields. Besides the field
content of the GL gravity theory is independent of the spacetime dimension
avoiding excessive number of terms in higher dimensions.

The results obtained here, along with the ones presented in \cite{CDIMR, CMR}%
, could be useful in order to construct a supersymmetric extension of the
pure Lovelock theory. \ Furthermore, the same procedure could be applied in
other (super)gravities in order to establish explicit relations between
non-trivial (super)gravity actions. In particular it would be interesting to
explore the existence of a configuration limit in order to derive the CJS
supergravity. In refs.~\cite{CR1, CFRS}, it has been suggested that the
Maxwell like superalgebras could be useful for such task.

Additionally, there are interesting features of the Lovelock formalism which
deserve to be explored in our generalized Lovelock theory. Of particular
relevance in the $AdS/CFT$ context are the black hole solutions of the
Lovelock gravity \cite{CS, C1, C2}. On the other hand, various problems in
the Lovelock gravity can be solved exactly \cite{BD} leading to a particular
interest in the effect of higher-curvature terms in the holography context
\cite{BLMSY, CES}. Moreover, matter conformally-coupled to gravity can be
seen as an extension of the Lovelock gravity theory \cite{OR}. In this
model, interesting problems related to the black hole geometry can be solved
exactly \cite{GLOR} allowing to study Hawking-Page phase transitions \cite%
{GGO, GGGO}. Further interesting studies about the Lovelock gravity theory
can be found in refs.~\cite{CHP1, CHP2, CH1, CHPP, CH2}.

Finally, it would be worth exploring our generalization to the
quasi-topological gravity which consist in higher curvature gravity \cite%
{OR1, OR2, MR, MPS, CFGO, CGHO, GN, DHM, LLL}. The field equations of such
theory reduce intriguingly to second order differential equations for
sphericaly symmetric spacetimes and have exact solutions similar to the
Lovelock ones. \ Such interesting behavior is not unique but appears in a
bigger family of theories that contains the Lovelock and the
quasi-topological theories, as well as the recent Einsteinian cubic gravity
theory \cite{BC1, HM, BC2} as particular examples \cite{HKM, BC3, AHMM}.

\section{Acknowledgment}

This work was supported by the Chilean FONDECYT Projects No. 3170437 (P.C.)
and No. 3170438 (E.R.). The authors wish to thank N. Merino and R. Durka for
enlightening discussions and comments.

\end{document}